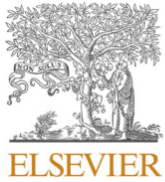
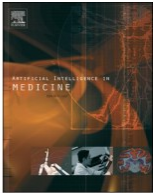

# Electrical and Mechanical Modeling of Uterine Contractions Analysis Using Connectivity Methods and Graph Theory

Kamil Bader El Dine[1,3], Noujoud Nader[2,3], Mohamad Khalil[3], Catherine Marque[1]

[1]CNRS UMR 7338, BMBI Sorbonne University, Université de technologie de Compiègne, Compiègne, France

[2]Louisiana State University, Baton Rouge, LA, USA

[3]Faculty of engineering, Azm center for research in biotechnology, Lebanese University

ABSTRACT

Premature delivery is a primary cause of fetal death and morbidity. Therefore, improving the prediction and treatment of preterm contractions is crucial. The electrohysterographic (EHG) signal measures the electrical activity that controls uterine contraction. Analyzing the features of the EHG signal can provide valuable information for labor detection. In this paper, we propose a new framework using simulated EHG signals to identify features sensitive to uterine connectivity. We focus on EHG signal propagation during delivery, recorded by multiple electrodes. We simulated EHG signals in different groups to determine which connectivity methods and graph parameters best represent the two main factors driving uterine synchronization: short-distance propagation (via electrical diffusion, ED) and long-distance synchronization (via mechanotransduction, EDM). Using the uterine model, signals were first simulated using just electrical diffusion by modifying the tissue resistance; second, signals were simulated using ED and mechanotransduction by holding the tissue resistance constant and varying the model parameters that affect mechanotransduction. We used the bipolar technique to construct our simulated EHGs by modeling a matrix of 16 surface electrodes (arranged in 4x4 matrix) placed on the abdomen of the pregnant woman. Our results show that even a simplified electromechanical model can be useful for monitoring uterine synchronization using simulated EHG signals. The differences seen between the selection performed by Fscore on real and simulated EHG signals shows that when employing the mean function, the best features are H2(Str), FW_h2 alone, and in combination with PR, BC, and CC. The best characteristics that demonstrate a shift in the mechanotransduction process are H2 alone or in combination with Str, R2(PR), and ICOH(Str). The best characteristics that demonstrate a shift in electrical diffusion are H2 alone and in combination with Eff, PR, and BC.

# 1. Introduction

Preterm labor (PL) is described as labor that happens before 37 weeks of pregnancy, whereas a normal pregnancy lasts 37 to 42 weeks, beginning from the first day of the last menstrual period [1]. Prematurity is the primary cause of newborn morbidity and mortality. The complications that occur because of prematurity are reported to be responsible for about one million newborn deaths globally each year [2]. Preterm births (PTB) account for 5-18% of all deliveries [3], with spontaneous preterm labor and preterm labor rupture of membranes accounting for 45% and 25%, respectively, of all preterm births [1].

During the latter half of the twentieth century, there was an alarming increase in the prevalence of preterm birth in the United States and other nations across the world [4]. According to the World Health Organization, more than 15 million newborns are born prematurely each year, with almost 1 million dying as a result of complications [5]. Preterm birth rates in the European Union, for example, range between 5 and 10%, while in the United States alone, the rate of preterm birth in 2014 was between 12 and 13% [6]. Multiple reasons were deemed to have contributed to the increased preterm birth rate including a higher average of the mother's age, more frequent use of assisted reproductive technologies, a rise in non-infertility-related multiple gestations, and higher rates of preterm inductions and cesarean sections. [7].

In addition, even if a preterm newborn survives, he/she may encounter a number of significant challenges, including breathing difficulties and eyesight impairments caused by undeveloped organs [8]. Furthermore, preterm labor may have a negative impact on maternal well-being because of the mother's perspective of her infant. In addition, the healthcare expenses of premature labor place a financial strain on both society and families, since the cost of such treatment is five to 10 times that of a term birth. As a result, early identification of premature labor, together with effective medical care to avert this phenomenon, is critical for enhancing newborn survival, the mother's mental health, and decreasing financial costs. One promising approach in this area is the electrohysterogram (EHG).

The electrohysterogram (EHG) is one of the potential approaches for diagnosing PL. EHG measures the electrical activity of the uterus on the mother's abdomen. EHG is advocated since it is inexpensive and needs basic equipment to capture uterine activity noninvasively [9]. EHG represents the electrical activity generated by active uterine muscle cells, along with the noise associated with corrupted electrical and mechanical operations. The EHG analysis has been demonstrated to be one of the most promising tools for monitoring uterine contraction efficiency throughout pregnancy [10].

Two physiological phenomena govern the efficiency of uterine contractions and the transition from pregnancy to labor: the excitability of uterine cells and the synchronization of the entire uterus. This synchronization arises from increased connectivity between myometrial cells, facilitated by the emergence of gap junctions and local diffusion, alongside enhanced long-distance synchronization via mechanotransduction processes [11] [12]. Consequently, numerous studies have concentrated on assessing the propagation or synchronization of uterine electrical activity, with EHGs proving highly effective in labor and pregnancy contractions classification [13] [12]. The nonlinear correlation coefficient (H2) has been instrumental in estimating correlations among 16 EHG signals recorded via a 4x4 electrode matrix, revealing significant distinctions in the characteristics of labor versus pregnancy EHG contractions [13]. Additionally, a previous study [14] revealed that multichannel EHG synchronization measures rose considerably closer to delivery. These findings imply that EHG synchronization analysis offers a new sensitive metric for detecting approaching labor, which could be utilized to improve preterm birth prediction and better understand uterine electrical activity dynamics. Moreover, EHG has been utilized to quantify the rate of electrical activity propagation, termed conduction velocity (CV) [15]. A combined analysis of peak frequency (PF) and propagation velocity (PV) has shown superior performance in discriminating between labor and non-labor EHGs [15].

On the other hand, a novel approach was presented, incorporating graph theory analysis along with connectivity methods to examine the association amid uterine electrical activities [16]. This method aimed to utilize graph parameters to describe the changes in uterine connectivity during the transition from pregnancy to labor. Subsequently, it aimed to distinguish contractions between those occurring during pregnancy and labor. After estimating the connectivity between EHG signals, the connectivity matrices were treated as graphs consisting of a set of nodes (representing electrodes) linked by edges (representing connectivity values between electrodes). The results of this study revealed an increase in EHG connectivity from the pregnancy stage to that of the labor [17] [18] [19]. This analysis has been also studied on simulated EHG signals [19]. Subsequently, the latter approach has been combined with Artificial Neural Networks. Herein, the parameters derived from connectivity and graph parameters are employed to improve the classification between labor and pregnancy using neural networks [9], [20].

However, the aim of this study is to analyze all the previously used parameters (connectivity methods with or without graph metrics) [20]. on .

simulated signals in order to select the features that are most sensitive to the electrical diffusion and the mechanotransduction process. This analysis allows us to identify the features that could be of interest to characterize uterine synchronization, and therefore the contraction efficiency. This paper is organized as follows. The process of data simulation and methods we employed, including a description of the uterine simulation model, are described in Section 2. The results of our investigations are presented in Section 3. Finally, the discussion and the concluding remarks of this work are included in Section 4.

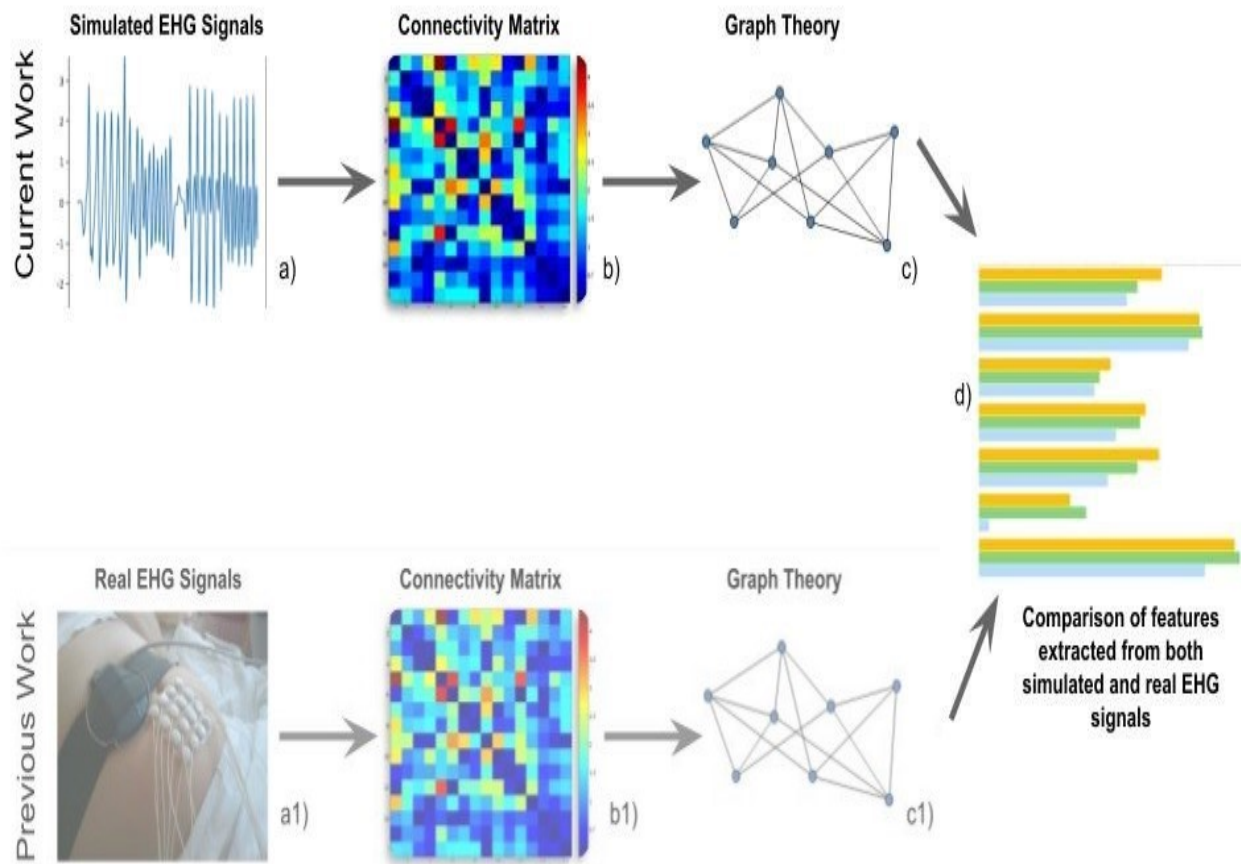

*Figure 1. Implementing structure. (a) Simulated EHG signals. (b) The Connectivity Matrix. (c) Graph Theory presentation (d) Compare the Simulated EHG signals results with the previous Real EHG signals results [20].*

## 2. MATERIALS AND METHODS

### 2.1 Structure of the Investigation

Figure 1 demonstrates the complete pipeline of this work. The first step represents the Simulated EHG signals (Figure 1a); then the connectivity matrix is calculated from the whole simulated signals (Figure 1b); this computed connectivity matrix is then transformed into a graph from which we extract measures (Figure 1c). Additionally, the recorded uterine EHGs are obtained by using a grid of 4x4 electrodes (Figure 1.a1), then the connectivity matrix is calculated using different connectivity methods from the whole signals (monopolar and denoised) (Figure 1.b1). Graph parameters are then extracted from these connectivity matrices for every approach (Figure 1.c1). In the final step, we compared the results with the previous results [20] retrieved from the real signals (Figure 1d).

In the following section, we will first explain the uterine contraction model, then discuss the data simulation process, followed by the connectivity method and graph theory. At last, we will examine the impact of various model parameters on the EHG characteristics by studying the two categories of simulated signals: ED and EDM.

## 2.2. Uterine Contraction Model

The measurement of the EHG signal on the mother's abdomen is influenced by various physiological phenomena occurring within the uterus. Understanding the connections between these physiological phenomena and EHG, as well as their significance, is crucial for comprehending uterine contractions and the potential for EHG-based monitoring. Modeling provides a promising avenue for studying these phenomena, offering the ability to simulate diverse configurations of physiological parameters. Successfully creating a model that accurately represents the uterus requires numerous steps. In this context, numerous modeling efforts have been undertaken to accurately replicate uterine physiology and simulate uterine contractions.

A recent method was introduced to enhance the understanding of uterine contractions and their correlation with EHG signals [21]. This approach utilizes numerical modeling to simulate the multi-physics and multi-scale phenomena inherent in uterine contractions. The model has been progressively refined to enhance accuracy and realism, incorporating electrical, chemical, and mechanical phenomena across various scales. This study focuses on further refining the model, starting with the integration of realistic geometry and the inclusion of the mechanotransduction phenomenon, which is currently facilitated by a simplified mechanical model.

## 2.3. Data simulation

The uterine muscle is modeled as a hyperplastic material whose active stress depends on the electrical diffusion. Aslanidi et al [22] developed a model taking into account the real geometry of a uterus (from MRI images) and modeled the propagation of electrical activity on this geometry. Sharifimadj et al [23] developed a uterine contraction model taking into account electrical, mechanical, and chemical phenomena. The electrical phenomena are modeled using a cellular excitability model of FitzHugh-Nagumo type which generates action potentials. Calcium concentration is calculated from action potentials using the model developed by Bursztyn [24]. The calcium concentration makes it possible to calculate the state of contraction of the cell, or the proportion of myosin and actin bound. The mechanical behavior is then determined from a model proposed by Sharifimadj et al [25], which has just distorted fibers (composed of contractile elements) longitudinal and circumferential in function of the proportion of actin and myosin bound in each contractile element.

The aim of this study is to compare the performance of different features used as input for the classification problem using simulated EHG signals. Specifically, we assess connectivity methods alone versus connectivity combined with graph parameters, while also representing the evolution of the electrical diffusion or the mechanotransduction process, we used simulated EHGs signals using a uterine simulation model.

This model is split into many sub-models that have been built to represent the process of mechanotransduction, which Young suggested as a new theory for understanding uterine activity synchronization during labor [12].

The first sub-model generates action potentials (APs) from ion exchanges across the cell membrane (Hodgkin-Huxley method) [26]. It also calculates the calcium concentration in each cell by simulating ionic action at the cellular level. The mechanical contraction model, which follows, utilizes this concentration as an input variable to calculate the force generated by each cell in relation to its electrical activity. These forces are then employed to compute the displacement of each node by the two sub-models that follow [27].

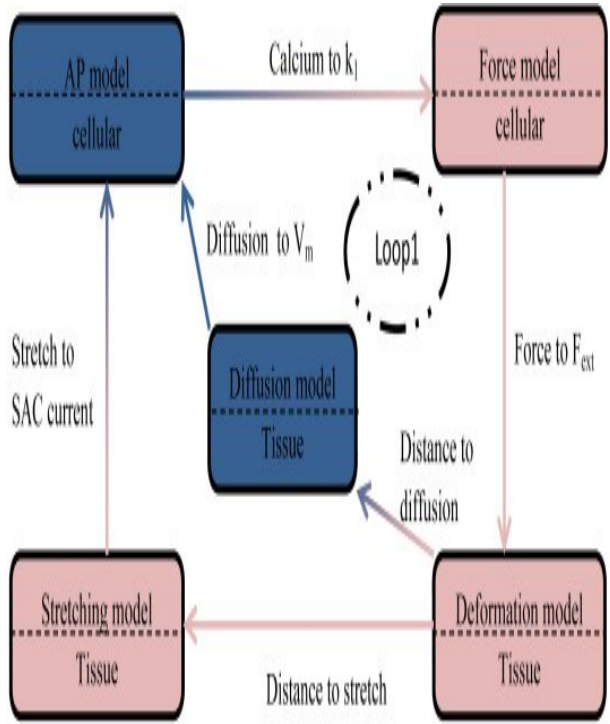

*Figure 32. Diagram of the uterine muscle model. The blue boxes represent the electrical models and the red boxes the mechanical ones. Notice that the arrow respects the color change when going from the electrical to the mechanical model (and vice-versa) [25].*

Based on these displacements, the model geometry is updated, and the stretches of each cell are determined. These strains affect the opening of stretching-sensitive ion channels, which are then incorporated into the cellular electrical sub-model. As a result, the stretched cells' electrical activity changes, resulting in new calcium concentrations, new forces, and a new phase in the simulation process connecting the multiple sub-models [28].

The simulated EHGs are derived by integrating the APs generated by each active cell, with the assistance of two additional sub-models: one representing the abdominal conducting volume (muscle, fat, and skin), and the other representing the electrodes.

Figure 2 illustrates the electro-mechanical component of the proposed model, providing an inclusive view of the uterine muscle model where the blue boxes represent the electrical models and the red boxes the mechanical ones. While Figure 3 complements this understanding by presenting a schematic representation of the conducting volume and electrode models.

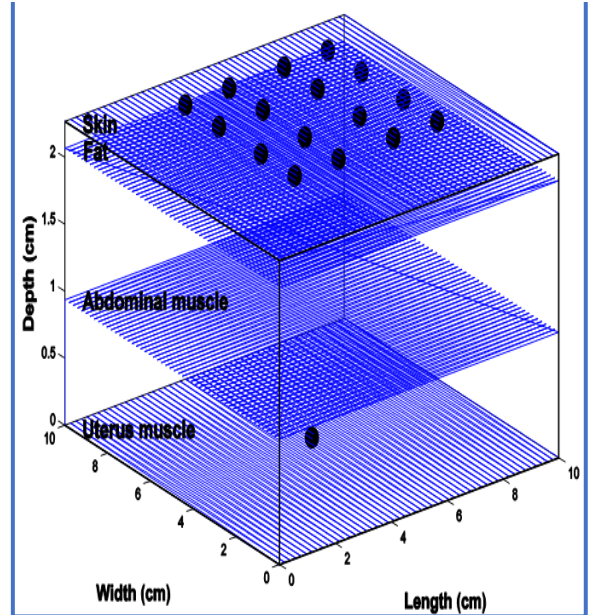

*Figure 23. Schematic representation of the conducting volume and of the electrodes [28]*

We used this uterine model to generate simulated signals used in this study.

We first generated signals in two classes to examine the influence of electrical and mechanotransduction model parameters on uterine synchronization. First, signals simulated with electrical diffusion alone (ED, group 1) by modifying tissue resistance; subsequently, signals simulated with electrical diffusion and mechanotransduction (EDM, group 2) by holding tissue resistance constant and varying the various model parameters that impact mechanotransduction. Mechanotransduction alone couldn't be examined because of the uterine tissue stretching; hence, mechanotransduction requires a certain quantity of electrical diffusion. We intend to use this study to determine the optimal features (connectivity alone, connectivity combined with graph metrics) that will allow us to track changes in EHG characteristics caused by altering model parameters [28].

Figure 4.a shows an example of a signal from group 1 (ED), whereas Figures 4.b, 4.c, 4.d, 4.e, and 4.f show signal samples from group 2 (EDM) with different values for the different parameters (Beta_sig($\sigma$), Current_Na_etirement (ICES_Na), Lambda_sig ($\lambda$), SACCH_current (ICES_Ca), and SACCH_nbmax(nbCES)). These parameters are shown in Table 1.

Beta_sig($\sigma$), where $\sigma$ is the stretch-sensitive channels (SSC) sigmoid shift. Current_Na_etirement (ICES_Na) where ICES_Na is the current of the ion current of sodium SSC. Lambda_sig (λ) where λ is the slope of the sigmoid that controls the opening of the SSC (stress sensitive channels). SACCH_current (ICES_Ca) where ICES_Ca is the current of ion for calcium SSC. SACCH_nbmax(nbCES) where nbCES is the number of SSC per cell.

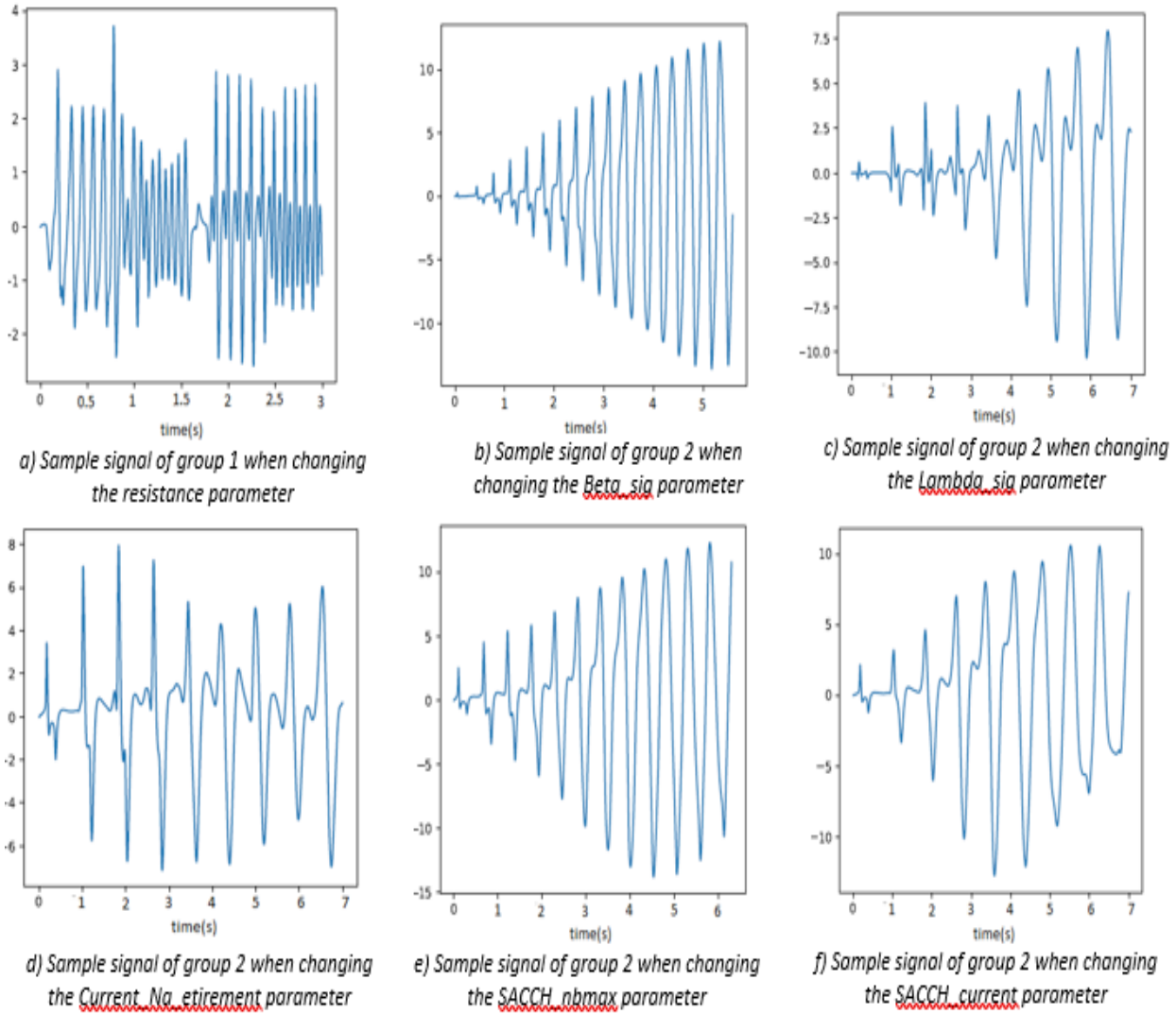

Figure 4. Different signals from group 1 and 2 with different values for the Beta_sig, Current_Na_etirement, Lambda_sig, SACCH_nbmax, and SACCH_current parameters

## 2.4 Connectivity Methods

We applied four connectivity measures within the 16 EHG signals, including nonlinear (H2) correlation coefficients, the classical linear (R2), the modified version of H2 called FW_h2, and the imaginary part of the coherence (ICOH). These measures have been demonstrated to be useful in previous studies [9], [17].

**The Nonlinear correlation (H2):** measures the nonlinear relationship between two variables X and Y. It is calculated by evaluating the value of X as a function of the value of Y from the two N-length signals X(t) and Y(t). For determining the value of Y given X, a nonlinear regression curve can be employed [29]. Subtracting the explained variance from the original yields the unexplained variance. H2, the nonlinear correlation value, represents the decrease in Y variance that may be derived by predicting the Y values from those of X, as H2 = (total variance - unexplained variance)/total variance, in accordance with the regression curve.

$$H2_{X/Y} = \frac{\sum_{k-1}^{N} Y(k)^2 - \sum_{k-1}^{N}(Y(k) - f(X_i))^2}{\sum_{k-1}^{N} Y(k)^2} \quad (1)$$

where f(Xi) is the nonlinear regression curve (linear piecewise approximation) and N is the length of the signal.

**The cross-correlation coefficient (R2):** indicates the strength of the time-domain linear correlation connection between two variables X and Y [30].

$$R2 = max_t \frac{cov^2(X(t), Y(t+\tau))}{var(X(t))var(Y(t+\tau))} \quad (2)$$

where cov and var are the covariance and variance among the two-time series X(t) and Y(t), respectively. ☐ reflects the change in time.

**Filtered Windowed H2 (FW_h2):** is a variation of the nonlinear correlation coefficient H2 [31]. This method involves filtering the Electrogastrogram (EHG) signal within a low-frequency band and then applying windowing techniques. It operates on the premise that EHG propagation is closely linked to its low-frequency bands (FWL: 0.1-0.3 Hz) [32]. Diab et al. [31] demonstrated that combining these two preprocessing stages resulted in the Filtered-Windowed-H2 (FW_h2) yielding superior results in classifying contractions between pregnancy and labor. Furthermore, FW_h2 exhibited a noticeable increase from pregnancy to labor phases.

**Imaginary part of coherence (Icoh):** Coherence is a measure extensively employed in the frequency domain to indicate the linkages between two signals X and Y as a function of frequency [33], where volume conduction has a direct impact on the true coherence value. Volume conduction takes place whenever electrical activity is captured and processed at a distance from its source, such as while monitoring abdominal EHGs. For this reason, novel solutions to this problem have been developed that are purely focused on the imagined component of coherence. The essential idea is that a zero-lag interaction of the real portions of the coherence function between signals indicates a spurious interaction. However, the imaginary component of the coherence function may disclose genuine interactions, indicating true signal correlation.

$$ICOH = \frac{|ImC_{XY}(f)|}{\sqrt{|C_{XX}(f)||C_{YY}(f)|}} \quad (3)$$

where the linear correlation between two signals X and X or Y as a function of frequency is represented by the C functions. Cxy is the imaginary part of the coherence between X and Y.

## 2.5. Graph Theory

The calculated connectivity matrices from the previous phase are then transformed into graphs. A graph is a mathematical abstract structure made up of vertices (V) or nodes and edges (E) that connect pairs of those vertices. In our work, nodes represent electrodes and edges represent computed connection (correlation) values [34]. In addition, five graph metrics were collected for each related graph: Strength (Str), Clustering Coefficient (CC), Efficiency (Eff), PageRank (PR), and Betweenness Centrality (BC).

**Strength (Str):** of a node reflects its relevance and connectivity in relation to other nodes in the network. The strength of a node is the total of the weights of the edges that link to it.

$$S_i = \sum_{j \epsilon N} w_{ij} \quad (4)$$

where $i$ and $j$ represent the $i_{th}$ and $j_{th}$ nodes, correspondingly. $N$ is the total number of nodes in the graph, and $w_{ij}$ denotes the connectivity value for the relationship between $i$ and $j$ [35].

**Clustering Coefficient (CC):** reflects the frequency with which nodes interact or connect to one another, and it depicts the degree to which a node's neighbors link to one another.

$$C_i = \frac{2t_i}{k_i(k_i - 1)} \quad (5)$$

where $i$ is the node, $t_i$ is the number of triangular connections between nodes, and $k_i(k_i-$

1) is the number of maximum possible edges in the network [17].

**Efficiency (Eff):** displays a proxy measure of network clustering properties [36]. It is the reciprocal of the shortest path between two nodes.

$$E = \frac{1}{N(N-1)} \sum_{i,j \in N, i \neq j} \frac{1}{d_{ij}} \quad (6)$$

where $i$ and $j$ denote the $i_{th}$ and $j_{th}$ nodes successively. The shortest path between two nodes $i$ and $j$ is represented by the value $d_{ij}$. N represents the total number of nodes in the network.

**PageRank (PR):** The PageRank algorithm continues to function on a network of nodes and edges. The relevance or effect of a node is estimated based on the contributions of other nodes linked to it, while taking the damping factor into consideration to reflect the possibility of following links or randomly switching to other

$$PR(u) = (1-d) + d \sum_{U \in Bu} \frac{PR(u)}{N_u} \quad (7)$$

nodes. The amount of links pointing to a certain node determines the PR value [37].

where $u$ signifies the node (electrode), $Nu$ the number of connections from $u$, and d the damping factor, which can range between 0 and 1. Bu is the set of pages that link to page u. These are the backlinks pointing to page u.

**Betweenness Centrality (BC):** The study of nodes that are frequently met on the shortest path between two other nodes [38]. As a result, betweenness centrality generates a relational value based on the local role of the node in relation to the nodes in between [39]. Nodes identified on a path between two other nodes oversee the flow of information between them, ranging from total control (when only one path exists between the two other nodes) to limited control (when many pathways exist between nodes [29]. It maintains track of how many times a node is placed on the shortest path between other nodes in the network. It assesses the researched node's ability to operate as a communication control point.

$$BC(v) = \sum_{s \neq v \neq t} \frac{\sigma st(v)}{\sigma_{st}} \quad (8)$$

where $\sigma st(v)$ is the number of shortest routes from $s$ to $t$ going via vertex $v$ and trail, and $st$ is the number of shortest paths from $s$ to $t$ [40].

## 2.6 ED and EDM study

As previously stated, the simulated signals were separated into two categories: ED and EDM. We investigated the impact of various model parameters on the EHG characteristics utilized in the connectivity analysis for each class. As a result, we first established the model parameters and the range of values assessed. Next, we adjusted the characteristics of the simulated EHGs, which were defined on the original EHGs.

### 2.6.1. Model parameters

In the first group (ED), the only model parameter that governs uterine synchronization via electrical diffusion is tissue resistance. In principle, when tissue resistance decreases, synchronization is expected to increase due to an easier diffusion [28]. As a consequence, considering the model's default resistance value is 40 [41], we examined the effect of changing this parameter using a range of values around this number. The examined values were in the range from 24 to 80, increasing in increments of 4.

On the other hand, five factors influence the mechanotransduction process in the second group (EDM, i.e. with long distance synchronization). For each of these factors, Table 1 shows the range of values selected to assess their influence on the EHG characteristics. We assume that the synchronization increases when the parameter value increases for all of the factors driving the mechanotransduction process as expected by the authors in [28].

In this framework, we performed 50 simulations for each scenario (model parameter selection). We then used the student test to determine whether or not the variations in feature values obtained with different model parameter values are significant [27].

Table 1. EDM Parameters

|  | Definition | Selected values |
|---|---|---|
| Lambda_sig (λ) | λ is the sigmoid slope that governs the SSC's opening (stress sensitive channels) | {3, 6, 9, 12, 15, 18, 21, 24, 27} |
| Beta_sig(σ) | σ is the SSC sigmoid shift | 1, 2, 3, 4, 5, 6, 7, 8, 9, and 10: [1-10] |
| SACCH_nbmax(nbCES) | nbCES is the number of SSC per cell | {20, 40, 60, 80, 100, 120, 140, 160, 180, 200} |
| Current_Na_etirement(ICES_Na) | ICES_Na is the ionic current for the sodium SSC (mA/cm2) | {0.005, 0.007, 0.009, 0.01, 0.03, 0.05, 0.07, 0.09, 0.11, 0.13} |
| SACCH_current(ICES_Ca) | ICES_Ca is the ionic current for the calcium SSC (mA/cm2) | {0.0007, 0.0009, 0.002, 0.004, 0.006, 0.008, 0.01, 0.013, 0.015, 0.017} |

**2.6.2 Frequency filter of FW_h2 method analyses**

We estimated the connectivity techniques for the simulated EHGs using H2, R2, and ICOH. However, we had to modify the utilized filter for FW_h2 to the spectral content of the simulated signals, which was not precisely the same as for the real EHG ones.

The real EHG is made up of two frequency components called FWL (Fast Wave Low, 0.1 to 0.3 Hz) and FWH (Fast Wave High, 0.3 to 2 Hz). The spread of uterine electrical activity is assumed to be more associated with FWL, whereas uterine excitability is thought to be more associated with FWH [42]. In this context, Terrien et al. [42] investigated the effect of filtering EHG signals into their various frequency components (low FWL and high FWH). Diab et al. [43] suggested a new feature, FW_h2, based on the nonlinear correlation approach, and demonstrated that filtering signals in the FWL band (0.1 to 0.3 Hz) increases the pregnancy/labor classification rate. This finding lends support to the concept that FWL is associated with uterine activity propagation and may represent the uterus' enhanced coordination during labor.

As a result, the filter in FW_h2 should be tuned to the FWL frequency range. We calculated the power spectral density (PSD) of these simulated signals for the different parameter values to analyze the filter tailored to the simulated EHGs. The PSD is used to characterize broadband random signals and were calculated using the Welch periodogram technique.

Figure 5 shows the PSD produced by adjusting the following parameters: beta_sig parameter, current_Na_etirement, lambda_sig, tissue resistance, SACCH_current, and SACCH_nbmax.

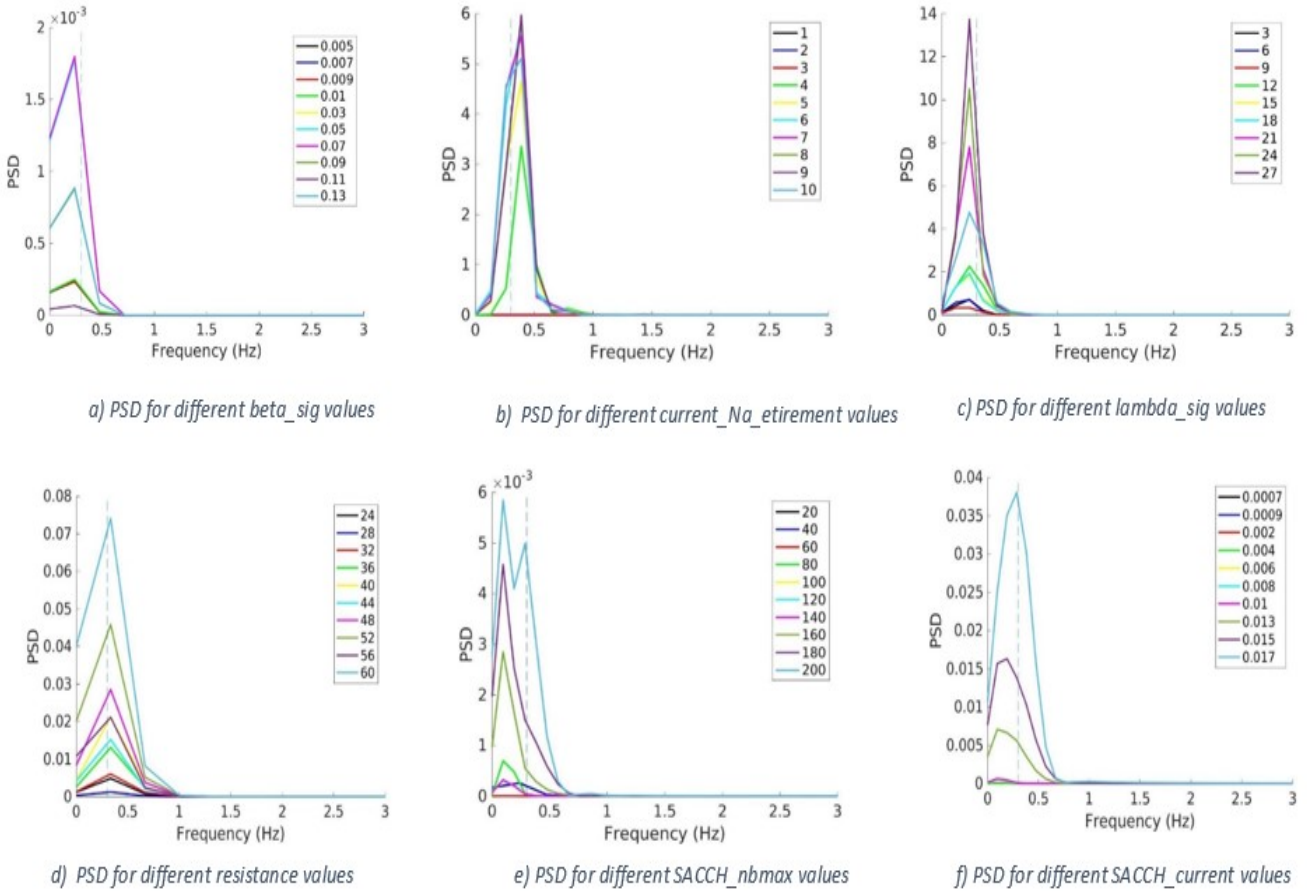

*Figure 5. PSDs of simulated EHGs*

As shown in Figure 5, the PSDs of simulated EHGs comprise primarily FWL at frequencies ranging from 0.1 to 0.7 Hz for all the parameters. Therefore, we will utilize a 0.1-0.7 Hz for FW_h2 filter applied to simulated EHGs to examine the influence of model parameters on the FW_h2 approach.

## 3. RESULTS

To assess the impact of parameter variations on synchronization, we conducted comparisons against a reference value. This process consists of comparing the findings with a reference value to investigate the impact of altering the different factors on synchronization. In each model parameter range, the reference value was specified as the first feature value defined for the first parameter value. The comparison was carried out by dividing each feature result by the reference. As a result, for all parameters, the value generated for the smallest value is assumed to be 1, and the variance is normalized. This approach allows us to assess whether the feature increases or decreases in response to the model parameter. If the value is less than one, the feature decreases, and if it is greater than one, the feature increases. Additionally, the positive slope indicates that the feature increases with the parameter, while a negative slope indicates the feature decreases as the parameter changes.

Figure 6 shows the results obtained for H2 when the tissue resistance was varied. The blue line represents the mean of all values of each resistance value, while the green line represents the median of all values of each resistance value. As evident, when the resistance value increases, the H2 value decreases. On the other hand, when the resistance of the tissue decreases, the synchronization is supposed to increase (easier diffusion).

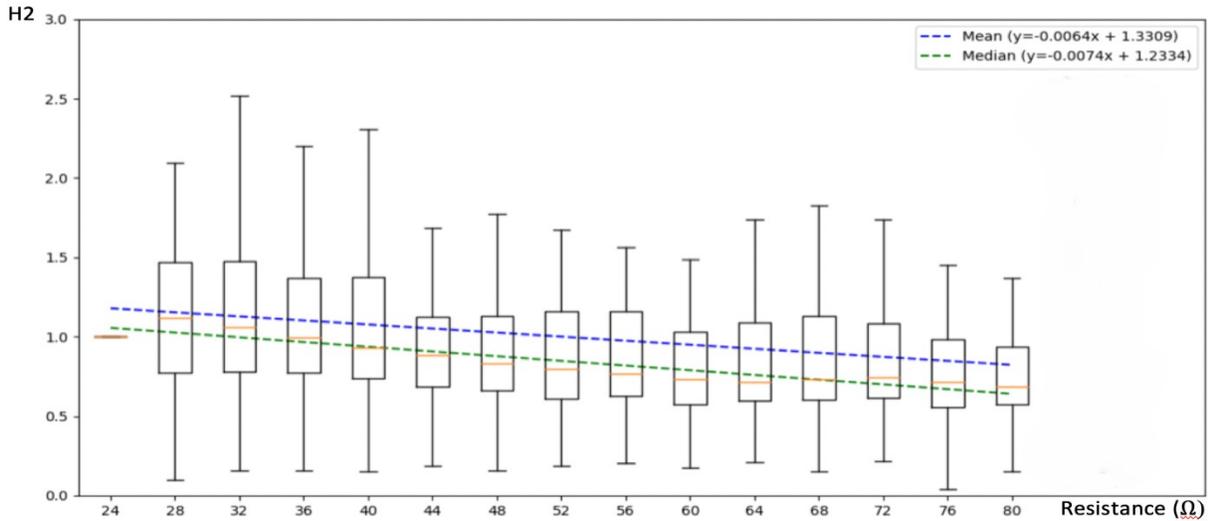

*Figure 6. Evolution of H2 function of the tissue resistance. Top right corner: equations of the linear regression computed from the mean and the variance of the feature values. Right column: results of the significative differences obtained for differ*

As previously indicated, our initial expectations centered around observing a decline in feature values, signifying a negative slope, for the resistance parameter within the ED group. Our anticipation was rooted in the hypothesis that alterations in this specific parameter would lead to a discernible decrease in the associated features. Concurrently, our predictions for the five parameters linked to mechanotransduction (EDM group) were aligned with an anticipated increase in connection methods as their respective values ascended, reflecting a positive slope. This hypothesis was grounded in the belief that heightened values of these parameters would correspond to an elevated level of mechanotransduction processes. Tables 2, 3, 4, and 5 present the slopes obtained for each connectivity method (R2, H2, ICOH, and FW_h2) with and without the incorporation of graph analysis metrics (Eff, BC, Str, CC, and PR). In these tables, the visual representation of these predictions, depicted in black, served as our baseline for expected outcomes. However, it is noteworthy that certain outcomes deviated from these projections, manifesting as unexpected results highlighted in red.

Table 2. R2 results for all the parameters

| | lambda_sig | | SACCH_nbmax | | beta_sig | | current_Na_etirement | | SACCH_current | | Resistance | |
|---|---|---|---|---|---|---|---|---|---|---|---|---|
| Method | Mean | Median | Mean | Median | Mean | Median | Mean | Median | Mean | Median | Mean | Median |
| R2 | -0.0102 | -0.0016 | -0.0123 | -0.0071 | 0.0353 | 0.0353 | -0.0002 | -0.0009 | -0.0073 | 0.0013 | -0.0009 | -0.0009 |
| R2(Eff) | 0.0033 | 0.0011 | 0.0001 | 0.001 | -0.0078 | -0.0039 | -0.0021 | -0.0003 | 0.0004 | 0.0004 | -0.0003 | -0.0004 |
| R2(BC) | -0.007 | -0.0041 | -0.0062 | -0.001 | 0.0143 | 0.0075 | 0.0009 | 0.0013 | -0.0047 | -0.0014 | -0.0008 | -0.0008 |
| R2(Str) | -0.0063 | -0.0036 | -0.0086 | -0.0012 | 0.0064 | 0.0063 | -0.0019 | 0.001 | -0.0071 | -0.0019 | -0.0014 | -0.0007 |
| R2(CC) | -0.0078 | -0.0047 | -0.0082 | -0.0043 | 0.0202 | 0.0124 | -0.001 | 0.0002 | -0.0035 | 0.0013 | -0.0013 | -0.0015 |
| R2(PR) | -0.0092 | -0.002 | 0.0006 | -0.0035 | 0.0073 | 0.0063 | 0.003 | -0.0002 | 0.0043 | -0.0001 | -0.0014 | -0.0019 |

Table 3. H2 results for all the parameters

| | lambda_sig | | SACCH_nbmax | | beta_sig | | current_Na_etirement | | SACCH_current | | Resistance | |
|---|---|---|---|---|---|---|---|---|---|---|---|---|
| Method | Mean | Median | Mean | Median | Mean | Median | Mean | Median | Mean | Median | Mean | Median |
| H2 | -0.0034 | -0.0034 | 0.0051 | 0.0049 | -0.0062 | -0.0084 | 0.0044 | 0.0039 | -0.0016 | -0.0038 | -0.0064 | -0.0074 |
| H2(Eff) | -0.0024 | -0.0021 | 0.0003 | 0.0038 | -0.0104 | -0.0064 | 0.0004 | 0.0026 | -0.0049 | -0.0028 | -0.0046 | -0.0042 |
| H2(BC) | -0.0038 | -0.0023 | -0.001 | 0.0001 | -0.0003 | -0.0005 | 0.0014 | 0.0011 | -0.0044 | -0.0048 | -0.0028 | -0.0028 |
| H2(Str) | 0.0255 | 0.0477 | 0.0189 | 0.0266 | 0.0038 | 0.001 | 0.0074 | -0.0145 | 0.0188 | 0.0228 | 0.0018 | 0.0001 |
| H2(CC) | -0.0019 | -0.0024 | -0.0015 | -0.0001 | -0.0036 | 0.0013 | -0.0014 | 0.0003 | -0.0011 | 0.0012 | -0.0004 | -0.0003 |
| H2(PR) | -0.0049 | -0.0039 | 0.0016 | 0.0013 | -0.0041 | -0.0035 | 0.0032 | 0.0036 | -0.0019 | -0.0019 | -0.0043 | -0.0044 |

Table 4. FW_h2 (filter: 0.1-0.7 Hz) results for all the parameters

| | lambda_sig | | SACCH_nbmax | | beta_sig | | current_Na_etirement | | SACCH_current | | Resistance | |
|---|---|---|---|---|---|---|---|---|---|---|---|---|
| Method | Mean | Median | Mean | Median | Mean | Median | Mean | Median | Mean | Median | Mean | Median |
| FW_h2 | 0.0017 | 0.0022 | 0.0018 | 0.0019 | -0.0003 | 0.0002 | 0.0008 | 0.0008 | 0.0009 | 0.0011 | 0.0006 | 0.0008 |
| FW_h2(Eff) | 0.0005 | 0.0006 | -0.007 | -0.0049 | -0.0031 | 0.0021 | 0.0001 | 0.002 | -0.0017 | 0.0011 | -0.0005 | -0.0004 |
| FW_h2(BC) | 0.0021 | 0.0027 | -0.0005 | 0.001 | -0.0006 | 0.0035 | 0.0002 | 0.0014 | -0.0017 | 0.0009 | -0.0017 | -0.0014 |
| FW_h2(Str) | 0.0015 | 0.0016 | -0.002 | 0.001 | -0.0032 | 0.0057 | -0.0024 | -0.0005 | -0.0027 | -0.0012 | -0.0001 | -0.0006 |
| FW_h2(CC) | 0.0007 | 0.0007 | 0.0006 | 0.0006 | 0.0018 | 0.0023 | 0.0006 | 0.0003 | 0.0004 | 0.0003 | -0.0037 | -0.0031 |
| FW_h2(PR) | 0.0089 | 0.0276 | 0.0121 | 0.0029 | 0.0221 | 0.0015 | 0.0014 | 0.0001 | -0.0016 | -0.0004 | -0.0003 | 0.0001 |

Table 5. ICOH results for all the parameters

|  | lambda_sig | | SACCH_nbmax | | beta_sig | | current_Na_etirement | | SACCH_current | | Resistance | |
|---|---|---|---|---|---|---|---|---|---|---|---|---|
| Method | Mean | Median | Mean | Median | Mean | Median | Mean | Median | Mean | Median | Mean | Median |
| ICOH | -0.0003 | 0.0019 | 0.0032 | 0.0054 | -0.0053 | 0.0005 | -0.0008 | -0.0014 | 0.031 | 0.0017 | 0.0001 | 0.0008 |
| ICOH (Eff) | 0.0002 | -0.0004 | -0.0021 | 0.0006 | -0.0063 | -0.0013 | -0.0031 | -0.0012 | -0.0023 | 0.0001 | -0.0007 | -0.0009 |
| ICOH (BC) | -0.0001 | -0.0005 | -0.0019 | 0.0013 | -0.0075 | -0.0081 | -0.0016 | 0.0004 | -0.0025 | 0.0003 | -0.0004 | -0.0002 |
| ICOH (Str) | -0.0029 | -0.0129 | 0.0168 | 0.0067 | -0.0053 | -0.0005 | 0.0147 | 0.032 | 0.0029 | 0.0064 | 0.002 | -0.0004 |
| ICOH (CC) | -0.0001 | -0.0005 | 0.0007 | 0.0013 | -0.0008 | -0.0008 | 0.0007 | 0.0005 | 0.0 | 0.0005 | 0.0003 | 0.0002 |
| ICOH (PR) | -0.0036 | -0.0039 | 0.0063 | 0.0003 | 0.0063 | 0.0008 | 0.0044 | 0.0011 | 0.0017 | 0.0024 | 0.0007 | 0.0001 |

In terms of electrical diffusion alone, the best result is clearly achieved with H2, which displays greater slopes, indicating a higher sensitivity to resistance fluctuation, for the majority of the graph parameters utilized.

Additionally, Table 6 shows the nine most sensitive aspects of the mechanotransduction process when computing the mean slope (first column) and median slope (second column).

As a point of reference, Table 6 also displays the nine best parameters previously chosen using the Fscore conducted by KB. El Dine et al. [44] and AUC (Area Under Curve) on real EHG signals that were retrieved from a previous study [17].

The features retrieved using FW_h2, either alone or in combination with graph parameters, are selected seven times out of the 18 best parameters chosen from real EHGs and nine times out of the 18 best parameters selected from simulated EHGs. Thus, FW_h2 (with or without graph parameters) appears to be important in characterizing the mechanotransduction process and uterine synchronization. On the other hand, R2 is the least effective technique, being chosen only three times, twice from real EHGs and once from simulated EHGs, and always coupled with a graph parameter.

Table 6. Best 9 features selected by the different methods used on real (Fscore and AUC) and simulated EHGs (Mean and median slopes). The features indicated in blue are the ones selected by Fscore

| **Simulated EHGs** | | *Real EHGs* [17] | |
|---|---|---|---|
| Simu_Mean | Simu_Med | *Real_Fscore* | *Real_AUC* |
| H2(Str) | FW_h2(BC) | *FW_h2 (Str)* | *ICOH (Str)* |
| FW_h2(PR) | H2(Str) | *ICOH (Str)* | *ICOH (Eff)* |
| FW_h2(CC) | ICOH(Str) | *ICOH (Eff)* | *ICOH (CC)* |
| R2(PR) | FW_h2(CC) | *ICOH (CC)* | *FW_h2 (Str)* |
| H2 | FW_h2 | *FW_h2 (BC)* | *H2 (PR)* |
| ICOH(PR) | FW_h2(Eff) | *H2 (BC)* | *FW_h2 (BC)* |
| FW_h2 | H2(Eff) | *FW_h2 (Eff)* | *H2 (BC)* |
| FW_h2(BC) | FW_h2(PR) | *R2 (Eff)* | *FW_h2 (Eff)* |
| ICOH(Str) | H2 | *FW_h2 (CC)* | R2 (BC) |

H2 (with or without graph parameters) is chosen more frequently using simulated EHGs than real EHGs. Furthermore, the characteristics retrieved from H2 appear to be primarily important when analyzing the electrical diffusion alone. This is consistent with the assumption that H2 represents the linear and non-linear correlation, which should be affected linearly by changes in tissue resistance.

Concerning the graph parameters, they appear to be most relevant when analyzing real EHGs, because only two connectivity techniques (H2 and FW_h2) are chosen, and only four times among the 18 best parameters chosen from simulated EHGs, and none of them are chosen for real EHGs (Table 7). As shown in Table 7, the strength graph parameter (Str) appears to be the best graph parameter (as proven by prior research on real EHGs [45]), followed by efficiency (EFF) and BC, validating the concept of this new graph parameter.

Table 7. Occurrence of each graph parameter among the best parameters selected from real and simulated EHGs.

| | None | Str | CC | Eff | PR | BC |
|---|---|---|---|---|---|---|
| **Simulated EHGs** | 4 | 4 | 2 | 2 | 4 | 2 |
| **Real EHGs** | 0 | 4 | 3 | 5 | 1 | 5 |
| *Sum* | *4* | *8* | *5* | *7* | *5* | *7* |

## 4. Discussion and Conclusion

We investigated the effect of model parameters that regulate uterine synchronization (electrical diffusion and mechanotransduction process) on uterine connectivity metrics retrieved from simulated EHG signals. We utilized a uterine simulation model to generate EHG signals into two groups: signals with electrical diffusion (ED) alone, where tissue resistance was varied, and signals with both ED and mechanotransduction (EDM), where many factors that govern this phenomenon were adjusted. We expected the connectivity measure to decrease in the ED group as resistance increased. On the other hand, we expected the connectivity measure to increase in the EMD group when the parameter values increase.

When employing the mean function, the best features are H2(Str), FW_h2 alone and combined with PR, BC, and CC as graph metrics. To track changes in mechanotransduction, the best features are H2 alone or combined with Str, R2(PR), and ICOH(Str). For electrical diffusion, the best features are H2 alone and with Eff, PR, and BC.

The FW_h2 method, with or without graph parameters, gave the best results for measuring uterine connection in both real and simulated signals, although it takes longer to run.

In this study, we demonstrated that the electromechanical model, despite its limitations, can successfully identify features suitable for monitoring uterine synchronization using simulated EHG signals. The differences observed between the feature selection results using Fscore on real and simulated signals may be due to the simplifications in our model. Unlike real EHGs, where multiple parameters can change simultaneously, our approach focused on analyzing the effect of one parameter at a time. This difference in approach likely explains some of the observed disparities between the real and simulated signals.

Another factor that may influence the results is the limited size of the electrode recording matrix (less than 10cm x 10cm), which does not allow for a thorough investigation of the mechanotransduction process associated with long-distance diffusion. As a result, the mechanotransduction process is not fully captured in the EHG signals collected or simulated in this study. A greater distance between electrodes should allow for more exact recording of this long-distance synchronization, allowing for a better understanding of this process.